\documentclass[fleqn,12pt,twoside]{article}
\usepackage{espcrc1}
\usepackage{epsfig,amssymb}

\usepackage[figuresright]{rotating}

     \newcommand{\pathnow}{}

\newcommand{\AmS}{{\protect\the\textfont2
  A\kern-.1667em\lower.5ex\hbox{M}\kern-.125emS}}

\title{Testing limits of  statistical hadronization}

\author{Johann Rafelski\address{Department of Physics,
           University of Arizona, Tucson, AZ 85721}%
        \thanks{Supported by grant DE-FG03-95ER40937 from 
               the U.S. Department of  Energy.}
  and
        Jean Letessier\address{Laboratoire de Physique 
                Th\'eorique et Hautes Energies\\
       Universit\'e Paris 7, 2 place Jussieu, F--75251 Cedex 05.}%
        \thanks{LPTHE, Univ.\,Paris 6 et 7 is:
Unit\'e mixte de Recherche du CNRS, UMR7589.}
}      
\begin{document}
\maketitle

\begin{abstract}
Much of the energy of the nuclei colliding at RHIC or SPS is converted into
final state hadronic particles. About a quarter of this energy is in 
baryons and antibaryons. There are nearly 10 strange quark pairs per
central rapidity participant. Do we really understand the hadronic particle 
yields? Do we need to introduce post-Fermi-model ideas such as chemical
 non-equilibrium in order to understand how a deconfined state hadronizes?
\end{abstract}

\section{STATISTICAL HADRONIZATION}
Statistical Fermi-Pomeranchuk  models have been used 
extensively to study particle yields and spectra 
since 1950  \cite{Fer50,Pom51}.  This approach was 
developed as a qualitative description
of the gross features of  particle production. It was originally
not meant to be  theoretically an accurate  picture, just a
phase space  estimate 
of what Fermi called an upper limit on particle production in 
strong interaction processes. In fact, Fermi defined the limitations
of the statistical approach stating three conditions:
1) exclusion of particles such as  photons  which have weak 
coupling to the  interacting system; 2)  absence of absolute
 chemical (abundance) equilibrium for many semi-weakly coupled particles; 3) 
requirement of  relative baryochemical equilibrium. At that time strangeness
was not yet discovered, and today we can add 3b) requirement of relative 
strangeness equilibrium. 

In the ensuing decade, another important feature of the strong interactions 
was discovered: the existence of numerous hadronic resonances. This property
of hadronic interactions 
poses a challenge for the statistical hadronization model
as the yield of particles is sensitive to 
the unidentified high mass hadron resonance states. 
We will discuss how this influences the expected particle yields 
in section \ref{secmassspec}. Our objective is to establish the 
magnitude of systematic theoretical error we have to expect given 
incomplete knowledge of the hadronic mass spectrum. A tacit 
assumption in our approach is  that  hadron-hadron
interaction is well described by the formation of resonant states,
and the remaining residual forces are negligible in comparison to the 
available energy content per particle. 

Understanding the hadro-chemistry, {\it i.e.,} of the composition of particles
produced, can finger-print the phase of matter which has undergone
statistical hadronization. Therefore, we will describe how chemical
parameters, the chemical potentials $\mu_i$ 
 and the phase space occupancy parameters $\gamma_i$,
allow to regulate the particle number. 
We show how the 
chemical non-equilibrium is  characterized in section \ref{nonchem}.
We then demonstrate, in section \ref{analysis},
that the {\it nearly complete particle production results} force upon us
the study of chemical non-equilibrium. We demonstrate in particular that a
precise description of all RHIC results obtained at 
$\sqrt{s_{\rm NN}}=130$\,GeV (RHIC-130) is possible within 
the scheme we propose. 

\section{EXPONENTIAL HADRON MASS SPECTRUM}
\label{secmassspec}
Fermi's statistical model provided the intellectual environment
for the theoretical understanding of 
the exponentially growing hadronic mass spectrum within 
the statistical bootstrap model invented by Rolf Hagedorn \cite{Hag65}.
Hagedorn recognized early on that the large number of different
hadronic states, and the increase in their number with their mass,
has great practical implications for the behavior of matter at high 
temperature, and can lead to the formation of a new phase of matter.

Within Fermi's statistical model
the number of heavy hadron resonances produced is exponentially 
suppressed with the ratio of the mass of the particle considered to
the temperature present. In spite of this, if
the number of different resonances per unit of mass, the  
mass spectrum $\rho(m)$, increases exponentially,
\begin{equation}\label{rhom}
\rho(m) \propto
     (m_0^2+m^2)^{\scriptstyle a/2}\,e^{\frac{m}{T_{\rm H}}}\,,
\end{equation}
there still would be a critical temperature of hadronic matter, 
corresponding to the (inverse) slope $T_{\rm H}$ 
of the exponential mass spectrum:
\begin{equation}
\ln {\cal Z}=\int d^3x  \int {d^3p\over (2\pi)^3} \int dm\,
       e^{-{\sqrt{m^2+p^2}\over T}}\, \rho(m)
       \Bigg\vert_{T\to T_{\rm H}}\hspace*{-0.6cm}\longrightarrow\infty, 
 \  \mbox{\ for\ } \ a > -\frac52 .
\end{equation}
Hagedorn called this  the boiling of hadronic matter \cite{Hag68}. 
Hot hadronic matter behaved at this point in the same way as in 
high temperature 1st order  phase transition~\cite{Hag80,Fio85}.

\begin{figure}[tb]
\vspace*{-4.1cm}
\centerline{
\psfig{width=11.cm,clip=,figure=\pathnow 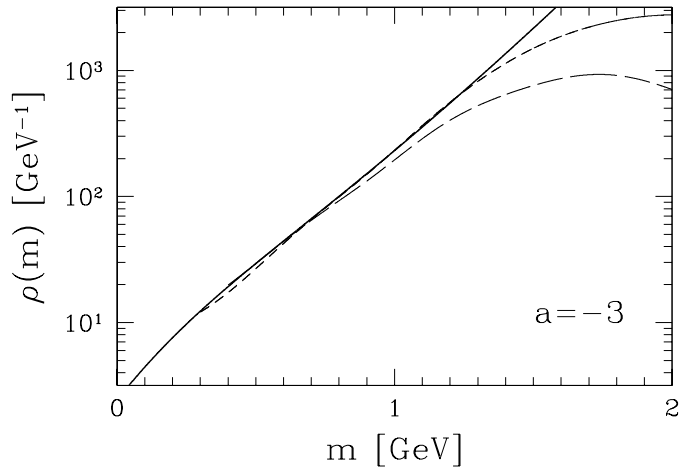}
}
\vspace*{-1.4cm}
\caption{ \small
Hadronic mass spectrum: solid line is the best-fit theoretical extrapolation,
dashed lines: experimental data (short dashed of 1996, long dashed of 1967).
\label{specm}
}
\end{figure}

There is strong empirical evidence that the experimental mass spectrum 
grows exponentially at least in the interval of interest to the
 study of particle production \cite{LetBook02,Bro00}. Smoothed 
mass spectrum based on 4627 states known in 1996 is shown 
as a short dashed line in figure \ref{specm}, while the 
1411 states of 1966, used by Hagedorn \cite{Hag67}, are depicted by long 
dashed line in figure \ref{specm}. The solid line is a fit to the 
experimental mass  spectrum  Eq.\,(\ref{rhom})
with the assumed value $a=-3$ which yields $T_{\rm H}=158$ MeV. 

Proposing  that the solid line describes 
reasonably the mass spectrum, we see that a significant number of
resonances remains to be identified for $m>1.4$ GeV (note the 
logarithmic scale in the figure \ref{specm}).  When a study of
particle yields is made, a significant 
systematic uncertainty derived from 
the lack of knowledge of hadronic mass spectrum can arise, in particular
so for models finding relatively high statistical hadronization 
temperatures. At a hadronization temperature of $T_h=145$ MeV, a study of 
the properties of the particle yields suggests that only about 5\% 
of pions are `missing'. However, at $T_h\simeq 175$ MeV \cite{PBM}, this 
increases to 35\%, and the high value of temperature 
considered poses a practical challenge, since the
Statistical Bootstrap model and the 
Lattice-QCD results for 2+1 flavors \cite{Ka02}
are converging to a critical temperature at $T_{\rm H}\lesssim 160$\,MeV.

\begin{figure}[tb]
\vspace*{-3.cm}
\centerline{\hspace*{0.5cm}
\psfig{width=9.5cm,clip=,figure=\pathnow 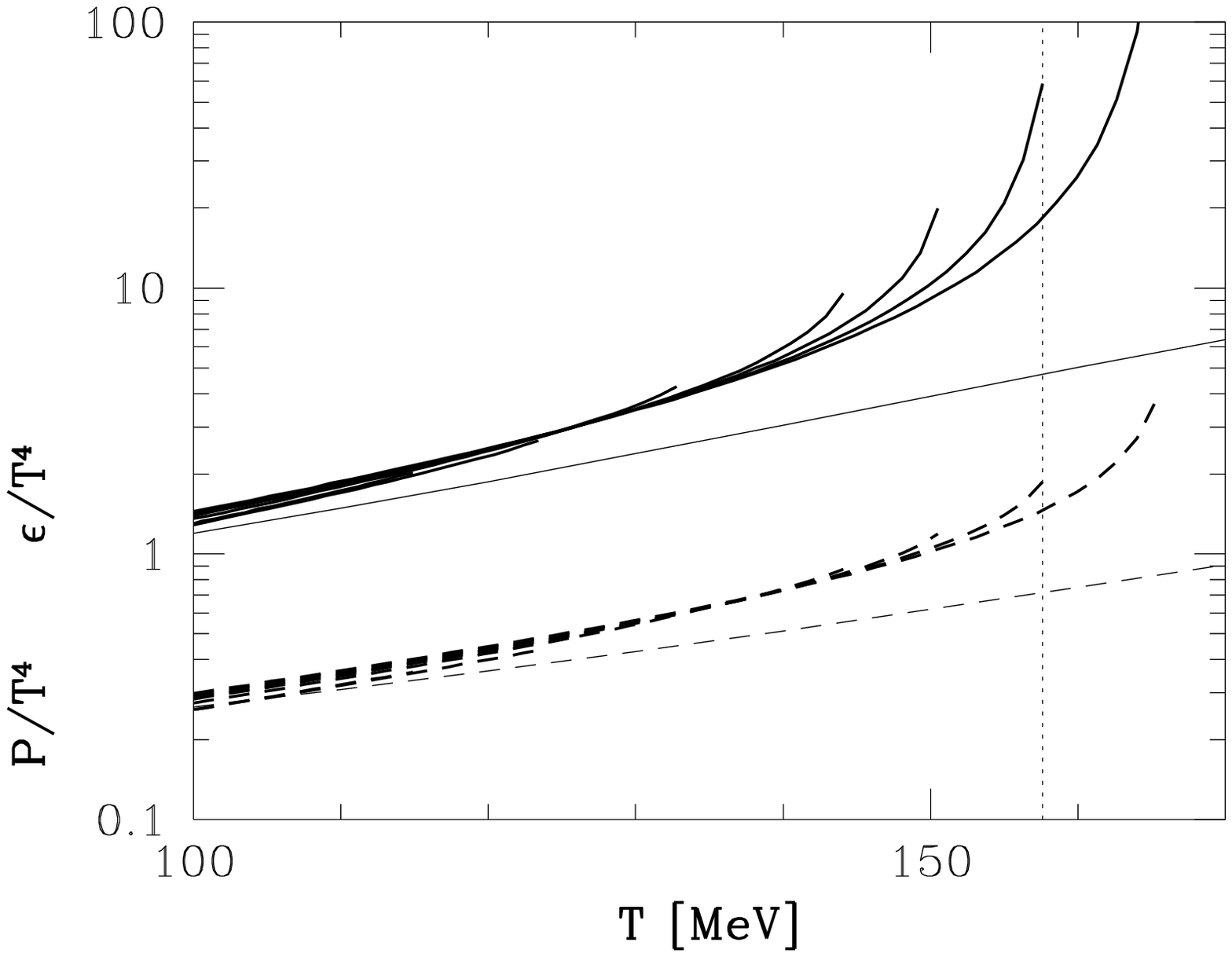}
\hspace*{-2cm}
\psfig{width=9.5cm,clip=,figure=\pathnow 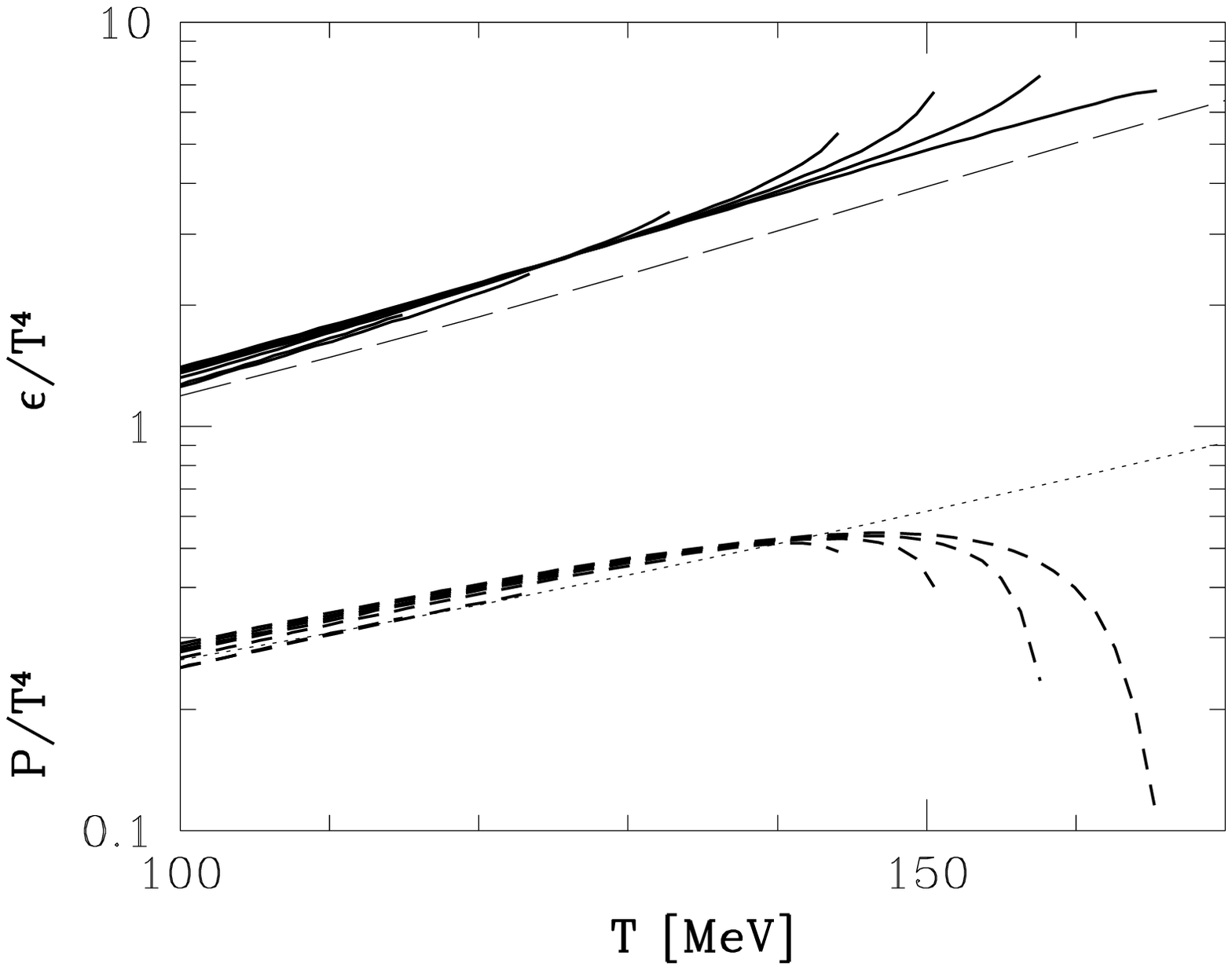}
}
\vspace*{-1.cm}
\caption{ \small
The energy density  $\varepsilon$ (solid lines) and pressure $P$
(dashed lines), both scaled with $T^4$. Left:  point hadrons with 
an exponential mass spectrum (thick lines) are compared to 
results using known hadrons (thin line). Right: hadron gas with finite 
volume correction, see text for more detail.
\label{pointgas}
}
\end{figure}

One way to assess the magnitude of the
systematic error  is to compare the properties of the hadron gas
using the mass spectrum of the
type given in Eq.\,(\ref{rhom}) with that studied using the currently 
known experimental mass spectrum. For a range of
values $a=-2.5 $  (most divergent curve in left frame in figure \ref{pointgas})
and $a=-7 $,  the energy density $\varepsilon$ (solid lines) and pressure $P$
(dashed lines), both scaled with $T^4$, are compared to the 
values obtained summing all known hadronic states 
(thin line in figure \ref{pointgas}). The differences as expected 
are dramatic at high temperature, near to the boiling 
point of hadronic matter.

It turns out that the problem is reduced considering the hadron proper
volume \cite{Hag80}. Reconsidering the energy density 
 assuming a proper energy density 
of hadrons as derived from the bag constant $\cal B$=$(190\,\mbox{MeV})^4$,
we find that the expected excess counting the energy density degrees of 
freedom remains at the level of 10\% as is seen 
in right  frame in figure \ref{pointgas}. While this allows to apply 
statistical hadronization method in study of particle abundance ratios, 
the influence of quantitatively uncertain finite size correction 
is very large,  implying in particular in the pion yield a considerable
systematic theoretical error.

\section{CHEMICAL NONEQUILIBRIUM }
\label{nonchem}
In general the fugacity $\Upsilon_i$ 
of each individual particle will comprise the 
two chemical factors associated with the two different chemical equilibria. 
For example, let us look at the nucleon, and the antinucleon fugacity and 
chemical potentials: 
\begin{equation}\label{UpsilonN}\label{sigmaN}
\Upsilon_N=\gamma_N e^{\mu_{\rm B}/T},\quad  
\Upsilon_{\overline{N}}=\gamma_N e^{-\mu_{\rm B}/T};\quad
\sigma_{N}\equiv \mu_{\rm B}+T\ln\gamma_N,  \quad
\sigma_{\overline{N}}\equiv -\mu_{\rm B}+T\ln\gamma_N.
\end{equation}
There is an obvious difference between the two chemical 
factors in Eq.\,(\ref{UpsilonN}):
the number of nucleon-antinucleon pairs is associated
with the value of $\gamma_N$ but not with\,$\mu_{\rm B}$. This can be seen looking at the first 
law of thermodynamics, in this context written as:
\begin{eqnarray}\label{1Emu}\nonumber
dE&=&-P\,dV+T\,dS+\sigma_N\,dN+\sigma_{\overline{N}}\,d\overline{N},\\
&=&
-P\,dV+T\,dS+\mu_{\rm B}(dN-d\overline{N})+T\ln\gamma_N(dN+d\overline{N}).
\end{eqnarray}
To obtain the second form, we have employed Eq.\,(\ref{sigmaN}). 
We see that $\mu_{\rm B}$ is the energy required to change 
the baryon number, 
$B\equiv N-\overline{N},$
 by one unit, while the
number of nucleon-antinucleon pairs, 
$2N_{\rm pair}\equiv  N+\overline{N},$
is related to $\gamma_N$. For $\gamma_N\to1$, the last term vanishes, at this
point small fluctuation in number of nucleon pairs does not influence the 
energy of the system, we have reached the  absolute baryochemical 
equilibrium. 

It is convenient to follow the quark flavor even in the study of hadron
yields, since this allows to keep the same notation across the 
phase boundary of quark matter and hadronic gas matter. We  use:
\begin{equation}\label{uds}\label{udq}
\lambda_u=e^{\mu_u/T},\quad \lambda_d=e^{\mu_d/T},\quad \lambda_s=e^{\mu_s/T},
\quad
\mu_q=\frac12 (\mu_u+\mu_d),\quad \lambda_q^2=\lambda_u\lambda_d.
\end{equation}
When the light flavors $u,d$ remain indistinguishable we introduce $\lambda_q$.

The relationship of baryochemical potential $\mu_{\rm B}$ and of
strangeness chemical  $\mu_{\rm S}$ to quark chemical potential is:
\begin{equation}\label{Ssrel}
\mu_{\rm B}=3\mu_q,\quad  \lambda_{\rm B}=\lambda_q^3,\quad
\quad \lambda_{\rm S}={\lambda_q\over\lambda_s},\quad
\mu_{\rm S}=\frac 1 3 \mu_{\rm B}-\mu_s.
\end{equation}
These relations arise considering that three quarks make a baryon, 
and remembering that strange quarks  carry {\it negative} strangeness 
and one third of baryon number. 

When using of fugacities which 
follow the valance quarks, $\lambda_i,\ i=u,d,s$,
hadronic particle yields can be easily  checked, and thus 
errors and omissions in a rather complex and large particle `zoo' minimized.
Consider, as an example, the ratio
$\overline{\Xi^-}(\bar d \bar s \bar s)/\Xi^-(dss)$.
Given the quark content and ignoring (at RHIC very small) 
isospin asymmetry, we find:
\begin{equation}\label{Xirat}
{\overline{\Xi^-}\over \Xi^-}=
{\lambda_s^{-2}\lambda_q^{-1}\over\lambda_s^{2}\lambda_q}
=\lambda_s^{-4}\lambda_q^{-2}
=e^{-4\,\mu_s/T}e^{-2\mu_q/T}
=e^{4\,\mu_{\rm S}/T}e^{-2\mu_{\rm B}/T},\quad
\mu_{\rm S}=\frac 12 \mu_{\rm B}+\frac T4 \ln {\overline{\Xi^-}\over \Xi^-}.
\end{equation}
Since all $\Xi$ resonances which contribute to this ratio are symmetric
for particles and antiparticles, and possible weak interaction 
feed from $\overline\Omega(\bar s \bar s \bar s) $, and respectively
 $\Omega(sss)$, are small, these expressions are actually nearly exact.
Thus we have inverted the relation and expressed $\mu_{\rm S}$ in terms of 
$\mu_{\rm B}$ and the cascade ratio. This allows the reader to check 
for correctness results presented elsewhere. Results shown in \cite{PBM}
appear inconsistent by 50\% while those in \cite{Bro02} are numerically 
consistent.
 
The important lesson to be drawn from Eq.\,(\ref{Xirat}) is that when we
compare in a ratio particle with antiparticle we `see' the chemical 
potentials. As we shall see below and use in section \ref{analysis}, 
the  chemical non-equilibrium parameters are
probed with {\it products} of yields of particles and antiparticles.

Near to chemical equilibrium,
we use  three non-equilibrium parameters,
$\gamma_s$, and $\gamma_u,\gamma_d$ or
equivalently just two parameters introducing, $\gamma_q=\sqrt{\gamma_u\gamma_d}$. 
In quark matter, these coefficients express the approach to 
the expected chemical equilibrium yield by the quark abundances. 
In a coalescence hadronization process, gluons fragment into quark pairs and
the net yields of quarks and antiquarks of all flavors 
 is redistributed among all  individual hadrons. Hence 
 even if there were no change of the quark pair number in hadronization, 
the values of $\gamma_u,\gamma_d,\gamma_s$
in hadron gas and quark matter must differ.  Moreover, the phase 
spaces have different size, and it is impossible in the rapidly evolving
fireballs to reequilibrate several quark flavors. Thus, 
we have to distinguish:   
\[\gamma_u^{\rm QGP},\gamma_d^{\rm QGP},\gamma_s^{\rm QGP},\quad
\mbox{from}\quad \gamma_u^{\rm HG},\gamma_d^{\rm HG},\gamma_s^{\rm HG}.\]

The lesson is that in an analysis of experimental data, we explore the 
properties and parameters of the hadron gas phase (HG). While these may be smooth 
across hadronization for the chemical potentials $\mu_i$, we expect the phase 
space occupancy parameters $\gamma_i$ to be quite different in the 
confined and deconfined phase.
We obtain the fugacities of all hadrons in terms of six parameters,
which implicitly carry the upper index `HG'.  

It is best to see
how this works looking at  typical examples:\\
a) baryons: protons $p(uud)$,
antiprotons $\bar p(\bar u\bar u\bar d)$, $\Lambda(uds)$, 
$\overline\Omega(\bar s\bar s\bar s)$, etc:
\begin{equation}\label{gambar}
\Upsilon_p=\gamma_u^2\gamma_d\,e^{2\mu_u+\mu_d},\ 
\Upsilon_{\bar p}=\gamma_u^2\gamma_d\,e^{-2\mu_u-\mu_d},\ 
\Upsilon_\Lambda=\gamma_u\gamma_d\gamma_s\,e^{\mu_u+\mu_d+\mu_s},\ 
\Upsilon_{\overline\Omega}=\gamma_s^3\,e^{-3\mu_s}.
\end{equation}
b) mesons:
$\pi^+(u\bar d)$, $\pi^-(\bar u d)$, $K^-(\bar u s)$, 
$\phi(\bar s s)$, etc:
\begin{equation}\label{gammes}
\Upsilon_{\pi^+}=\gamma_u\gamma_d\,e^{\mu_u-\mu_d},\quad
\Upsilon_{\pi^-}=\gamma_u\gamma_d\,e^{-\mu_u+\mu_d},\quad
\Upsilon_{K^-}=\gamma_u\gamma_s\,e^{-\mu_u+\mu_s},\quad
\Upsilon_{\phi}=\gamma_s^2.
\end{equation}
Note that in the products of particle $\Upsilon_{\rm P}$ and antiparticle 
$\Upsilon_{\rm A}$ fugacity the 
chemical potentials $\mu_i$ always 
cancel $\Upsilon_{\rm P}\Upsilon_{\rm A}=f(\gamma_i)$. 

In the statistical hadronization approach, we have the same value of $\Upsilon$ for
all hadrons with the same valance quark  content. For example
$\Upsilon_p=\Upsilon_{\Delta^+}$. Thus, we assume tacitly that the relative
population of heavier resonances is in chemical equilibrium with the
lighter states.  This means that statistical hadronization 
resolves the valance 
quark and antiquark distribution  and does not allow for the 
possibility that heavier resonances may simply not be populated.
This method, thus, is most suitable for 
a hadronizing quark matter fireball, and may miss important 
features of a hadron fireball which never entered the deconfined phase. 

\begin{figure}[tb]
\vspace*{-3cm}
\centerline{\hspace*{0.5cm}
\psfig{width=9.5cm,figure=\pathnow 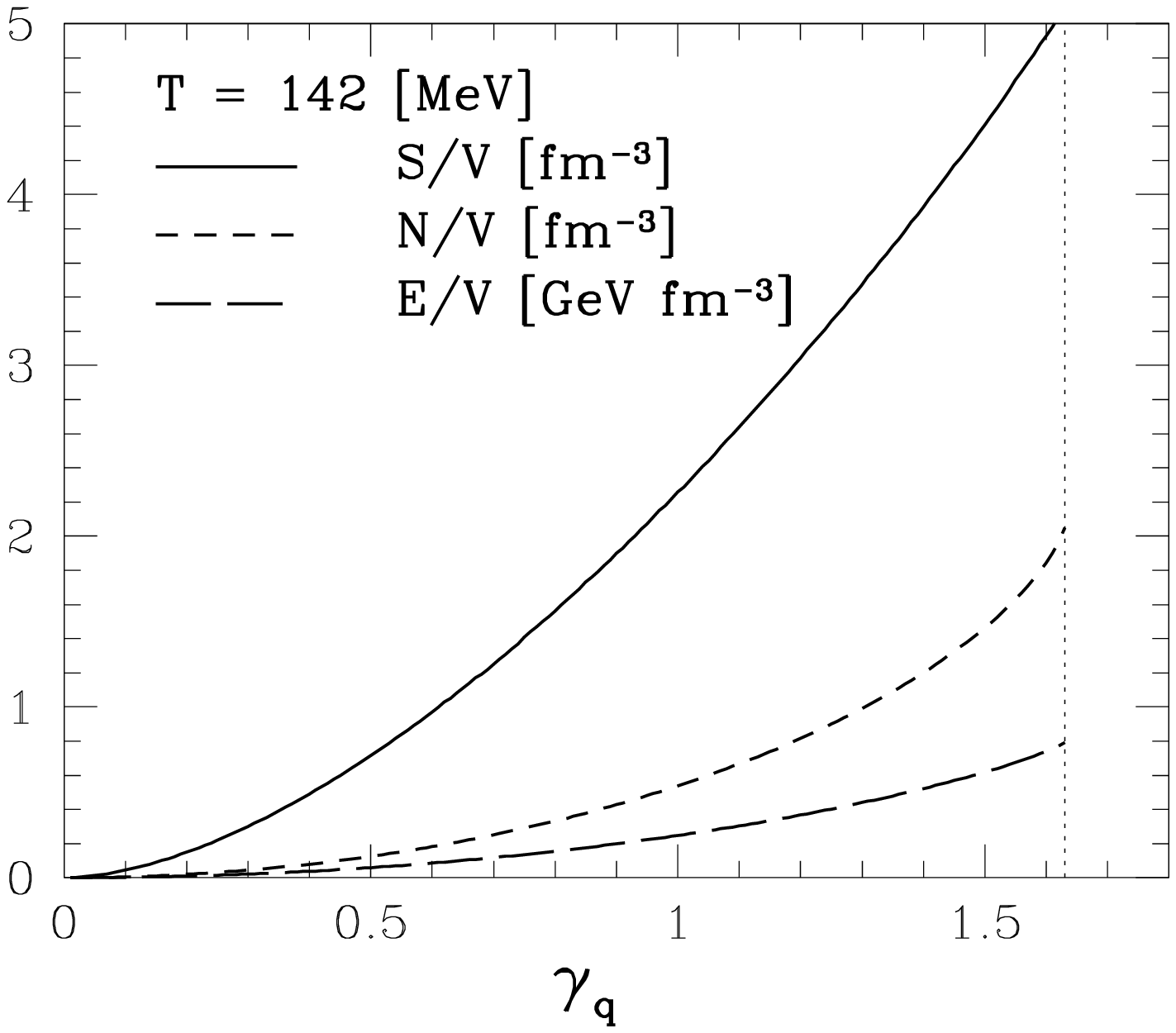}\hspace*{-2cm}
\psfig{width=9.5cm,figure=\pathnow 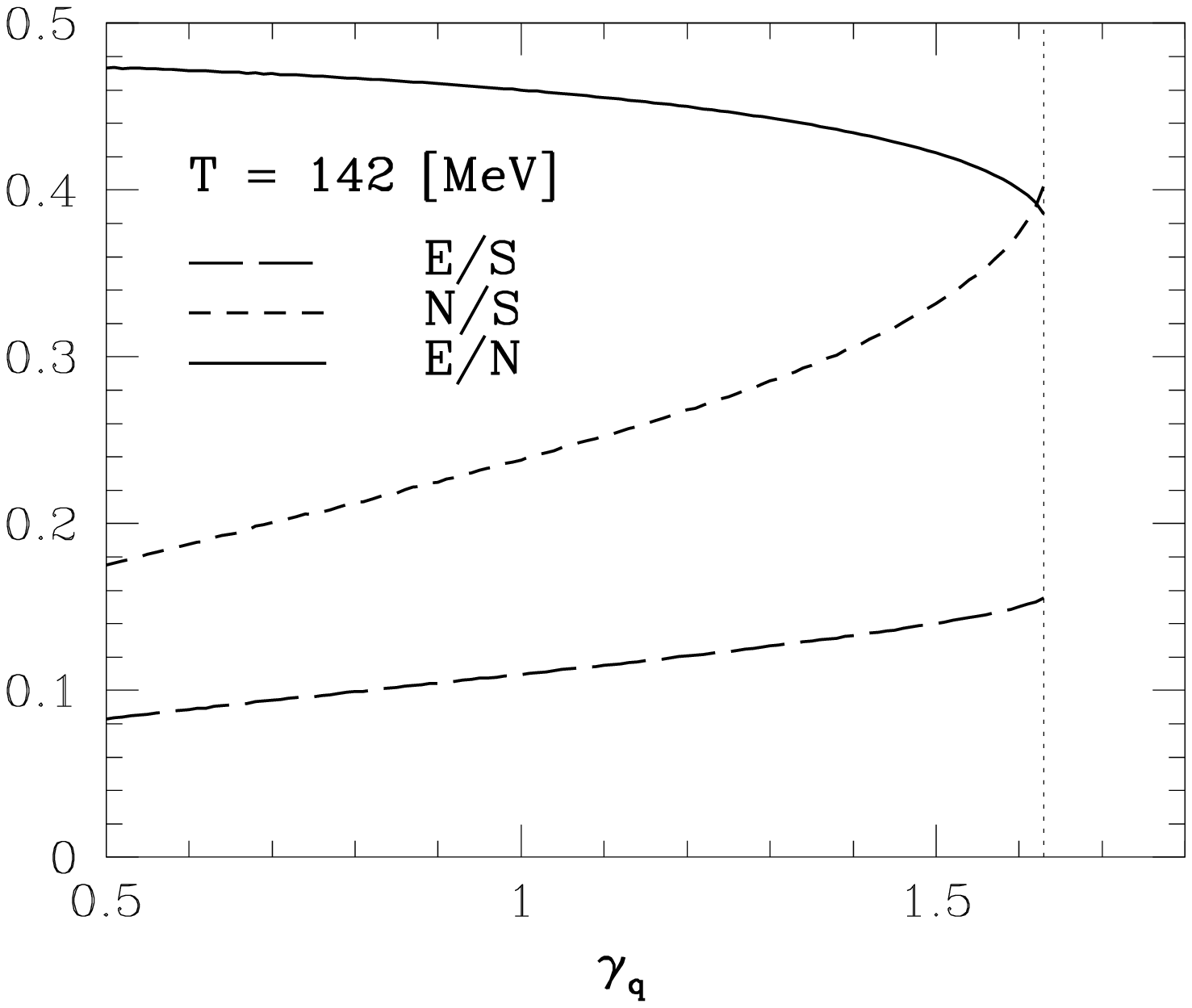}
}
\vspace*{-1.5cm}
\caption{ \small
Pion gas
 properties as function of chemical nonequilibrium parameter $\gamma_q$.
\label{piongas}
}
\end{figure}

The physical properties of the gas of hadrons 
with $\gamma_u,\gamma_d,\gamma_s\ne 1$ have certain highly 
desirable properties. Consider the properties of the pion gas
as function of $\gamma_q$, and in particular the entropy (B for Bosons and 
F for fermions):
\begin{equation}
{S}_{\rm B/F}=
 \int\!\frac{d^3\!p\, d^3\!x}{(2\pi\hbar)^3}\,
   \left[\pm(1\pm f)\ln(1\pm f)- 
             f\ln f\right]\,,\qquad 
f_\pi=\frac{1}{\gamma_{q}^{-2}e^{E_\pi\over T}-1}\,.
\end{equation}
Here, $ E_\pi=\sqrt{m_\pi^2+p^2}$. We see, in 
figure \ref{piongas}, that the entropy 
density is rising rapidly and it nearly doubles with $\gamma_q$ increasing
from the equilibrium value $\gamma_q=1$ toward condensation 
point $\gamma_q^2=e^{m_\pi/T}$. This high entropy content of the oversaturated
pion gas allows the 
sudden hadronization of QGP without reheating or inflation 
(volume expansion). A striking feature of the experimental 
data analysis which allows
for  $\gamma_q$ is the maximization of entropy density in pion gas,
{\it i.e.,} natural tendency toward the value $\gamma_q^2=e^{m_\pi/T}$.

\section{DATA ANALYSIS}
\label{analysis}
There are in our approach at most 6 chemical
parameters, and  hadronization temperature. 
Especially at RHIC,
we need not distinguish the light flavors $u,d$ and thus the number of 
statistical parameters is reduced by two. A further reduction is 
arrived at when we demand that strangeness balances anti-strangeness
locally, and finally  the requirement
that when entropy rich quark-gluon plasma hadronizes, the yields of 
hadrons are maximizing the entropy content of hadrons, yields
$\gamma_q^2=e^{m_\pi/T}$. Thus, we are in fact needing just two chemical
parameters which are usually $\lambda_q$ and $\gamma_s$.
Both at SPS \cite{Tor01} and at RHIC \cite{Bro01}, there is good evidence
for a single freeze-out to apply, in consistency with the sudden hadronization
hypothesis. In other words, one single temperature allows to understand 
both yields and spectra of hadrons. 

The computation of the particle yields is 
much simpler than for SPS. For central rapidity, we
have,  at RHIC-130,  approximate  
longitudinal scaling. Thus, we can act as if a series of
fireballs at all rapidities was present. Then, we simply
evaluate the full phase-space yields in order to obtain particle 
ratios. We do not include in our analysis trivial results
such as ${\pi}^+/{\pi}^-=1$. We also do not fit the results for
$\rm K^*$ and $\rm\overline{K^*}$ since the reconstructed yields depend on 
the degree of rescattering of resonance decay products. 

We first convince ourselves that the introduction of
 chemical non equilibrium in the study of hadron abundances is
necessary --- even though one could argue that if it is not
needed, the data fit will converge to chemical equilibrium 
condition. Therefore before we present
an overall fit of many particle yields, we present a short
but persuasive argument.

As noted, when we evaluate the product of
 particle and antiparticle yields, 
the chemical potentials $\mu_i$ cancel. Studying such products allows 
to focus only on $T,\gamma_q,\gamma_s$. 
We next identify  three ratios  involving
products of particles and antiparticles:
\begin{equation}\label{multiratio}
\sqrt{\frac{\Xi\overline\Xi}{\Lambda\overline\Lambda}}
     \propto \frac{\gamma_s}{\gamma_q}\  e^{-\frac{m_\Xi-m_Y}T};\quad
\sqrt{\frac{\Lambda\overline\Lambda}{p\bar p}}\propto 
    \frac{\gamma_s}{\gamma_q}\  e^{-\frac{m_\Lambda-m_N}T};\quad
\sqrt{\frac{K^-K^+}{\pi^-\pi^+}}\propto 
\frac{\gamma_s}{\gamma_q}\ e^{-\frac{m_K-m_\pi}T}.
\end{equation}
By comparing  mesons with mesons, 
and baryons with baryons we
reduce uncertainties about excluded volume.
 
The ratios considered in Eq.\,(\ref{multiratio}) have only two 
parameters $T$ and $\gamma_s/\gamma_q$. Thus, it is possible to 
present, in a two dimensional figure, how these three ratios 
behave, as is seen in figure \ref{GGT} for the experimental data
obtained at RHIC-130 (see table  \ref{RHIChad} below).  
Dashed area shows the allowed parameter 
domain. The cross in the figure is the result of global 
data analysis discusses below where the ratio of 
baryons to mesons, not used here, fixes the temperature.
While the kaon to pion ratio would tolerate within one standard deviation 
the chemical equilibrium, the baryon double ratios, and in 
particular the more strange cascades, are clearly demanding
a  value of $\gamma_s/\gamma_q>1$. 

We use the latest experimental results for the ratio
of hyperons to nucleons (yields of  $p,\bar p$ are weak decay feed 
corrected, $\Lambda,\overline\Lambda$ are uncorrected) \cite{Phe02}. 
This result is nearly by a factor {\it two}
different from the experimental 
data stated in \cite{Bro02}, and this reference also 
does not consider $\Xi,\overline\Xi$  results other than their ratio, 
and is therefore consistent with the
chemical equilibrium in the result of its analysis.

\begin{figure}[tb]
\vspace*{-3.8cm}
\centerline{
\psfig{width=11.7cm,figure=\pathnow 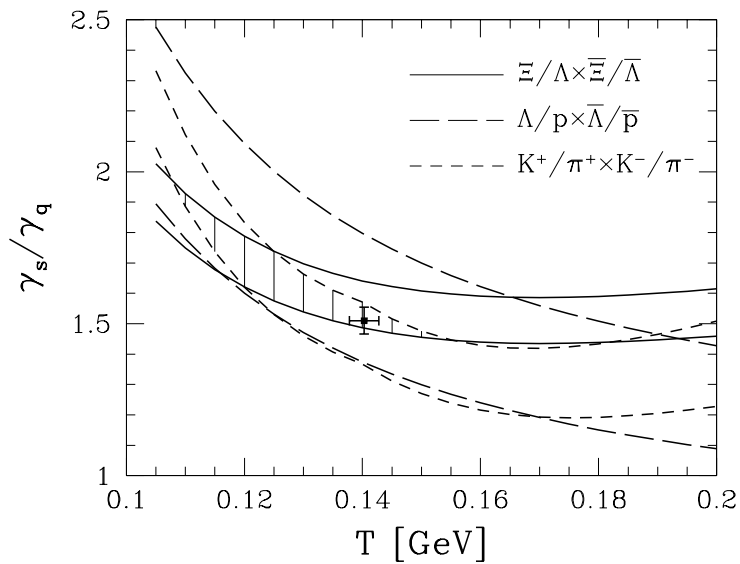}
}
\vspace*{-1.6cm}
\caption{ \small
Ratios of products of particle and antiparticle abundances in the 
$T$-$\gamma_s/\gamma_q$ plane. Ranges derived from data shown
in table \protect\ref{RHIChad}.
\label{GGT}
}
\end{figure}

\begin{table}[!p]
\caption {\label{RHIChad}\small\baselineskip=0.4cm
Central-rapidity hadron ratios at RHIC-$\sqrt{s_{\rm NN}}=130$ GeV.
From top to bottom: experimental results, 
fitted chemical parameters, the
physical properties of final state hadron phase-space,  and the fitting error. 
Columns: ratio considered, data value with reference, two non-equilibrium
fits,  and in the last column, 
the chemical equilibrium fit. 
The superscript $^*$ 
indicates quantities fixed and not fitted. 
The superscript $^\dagger$ indicates the error given is dominated by
theoretical considerations. Subscripts $\Xi,\Lambda$ mean that these values
include weak cascading given in heading of table. In parenthesis, we show  the contribution 
of the particular result to the total $\chi^2$.
}
\small
\begin{tabular}{lccccc}
\hline
  {\vphantom{$\frac AB$}}               &                 &      & 100\%\,$\Xi\to Y$&40\%\,$\Xi\to Y$ &40\%\,$\Xi\to Y$ \\
  {\vphantom{$\frac AB$}}               &Data             & Ref. & 40\% \ $Y\to N$  &40\%\,$Y\to N$   &40\%\,$Y\to N$  \\
\hline  
$ {\bar p}/p${\vphantom{$\frac AB$}}       &0.71 $\pm$ 0.06    &\cite{Phe02}&0.672(0.4) & 0.677(0.3)& 0.688(0.1)\\
${\overline\Lambda_\Xi}/{\Lambda_\Xi}$     &0.71 $\pm$ 0.04    &\cite{Sta02}&0.750(1.0) & 0.747(0.8)& 0.757(1.4)  \\
${\overline{\Xi}}/{\Xi}$                   &0.83 $\pm$ 0.08    &\cite{Cas02}&0.793(0.2) & 0.803(0.1)& 0.818(0.0) \\
$\rm{K^-}/{K^+}$                           &0.87 $\pm$ 0.07    &\cite{Phe01}& 0.925(0.6)& 0.924(0.6)& 0.933(0.8) \\
$\rm{K^-}/{{\pi}^\pm}$           & 0.15 $\pm$ 0.02$^\dagger$   &\cite{Phe01}& 0.156(0.1)& 0.157(0.1)& 0.150(0.0)  \\
$\rm{K^+}/{{\pi}^\pm}$           & 0.17 $\pm$ 0.02$^\dagger$   &\cite{Phe01}& 0.169(0.0)& 0.170(0.0)& 0.161(0.2) \\
${\Lambda_\Xi}/{h^-}$            & 0.059 $\pm$ 0.004$^\dagger$ &\cite{Sta02}& 0.057(0.6)& 0.049(6.7)& 0.047(9.6)  \\
${\overline\Lambda_\Xi}/{h^-}$   & 0.042 $\pm$ 0.004$^\dagger$ &\cite{Sta02}& 0.042(0.0)& 0.036(2.0)& 0.035(2.8) \\
${\Lambda_\Xi}/p$                          & 0.90 $\pm$ 0.12\  &\cite{Phe02}& 0.805(0.6)& 0.662(3.9)& 0.494(11.5) \\
${\overline\Lambda_\Xi}/\bar p$            & 0.93 $\pm$ 0.19\  &\cite{Phe02}& 0.899(0.0)& 0.731(1.1)& 0.543(4.1)  \\
$ \pi^\pm/p_\Lambda$                         &9.5 $\pm$ 2      &\cite{Phe01}& 9.4(0.0)  & 9.2(0.6)  & 7.4(27.7)  \\
$ \pi^\pm/{\bar p}_\Lambda$                  &13.4 $\pm$ 2.5   &\cite{Phe01}&13.7(0.1)  &13.3(0.0)  &10.6(9.6) \\
${\Xi^-}/{\pi}$                    &0.0088$\pm0.0008^\dagger$  &\cite{Cas02} &0.0092(0.2) & 0.0097(1.2) &  0.0069(5.8)  \\
${\Xi^-}/{h^-}$                    &0.0085$\pm$0.0015          &\cite{Cas02} &0.0076(0.3) & 0.0079(0.2) &  0.0056(3.8)  \\
${\overline{\Xi^-}}/{h^-}$         &0.0070$\pm$0.001           &\cite{Cas02} &0.0061(0.9) & 0.0064(0.4) &  0.0046(6.0)  \\
${\Xi^-}/{\Lambda}$                           &0.193$\pm$0.009 &\cite{Cas02} &0.190(0.1) & 0.189(0.2) &  0.132(45.4)  \\
${\overline{\Xi^-}}/{\overline\Lambda}$       &0.221$\pm$0.011 &\cite{Cas02} &0.207(1.6) & 0.206(1.9) &  0.144(48.4)  \\
${\Omega}/{\Xi^-}$                         & & &0.20 & 0.21 &  0.18 \\
${\overline{\Omega}}/{\overline{\Xi^-}}$   & & &0.22 & 0.23 &  0.20 \\
${\Omega}/{h^-}$                           &0.0012$\pm$0.0005  &\cite{Sui02}&0.0015(0.4) & 0.0016(0.7) &0.0010(0.13)   \\
${\overline{\Omega}}/{\Omega}$             &0.95$\pm$0.1       &\cite{Sui02} &0.87(0.7) & 0.88(0.5) &  0.90(0.3){\vphantom{$\frac AB$}} \\
%
${\phi}/K^-$                               &0.15$\pm$0.03      &\cite{Sta02PHI} &0.174(0.6)& 0.177(0.9) &  0.148(0.0)   \\
${\phi}/h^-$                               &0.021$\pm$0.001    &\cite{Sta02PHI} &0.022(1.3)&0.023(2.5)  &0.018(10.2)     \\
\hline  
$T${\vphantom{$\frac AB$}}     & &  &140.3 $\pm$ 1.1  & 142.5 $\pm$ 1.2    &165.8 $\pm$ 2.2   \\
$\gamma_{ q}^{\rm HG}$         & &  &1.64$^*$         & 1.63$^*$           &  1$^*$            \\
$\lambda_{ q}$                 & &  &1.0700 $\pm$ 0.0076& 1.0686 $\pm$ 0.0076 &1.0654 $\pm$ 0.0082\\
$\mu_{\rm B}$ [MeV]                  & &  &28.5             &28.4                &31.3                 \\
$\gamma_{ s}^{\rm HG}/\gamma_{ q}^{\rm HG}$ & &  &1.50 $\pm$ 0.04   &  1.48 $\pm$ 0.04    &  1$^*$               \\
$\lambda_{ s}$                 & &  &1.0243$^*$       &  1.0218$^*$        & 1.0186$^*$            \\
$\mu_{\rm S}$ [MeV]            & &  &6.1              &6.4                 &7.4                    \\
\hline{\vphantom{$\frac AB$}}
$E/b$ [{\small GeV}]    $\!\!$ & &  &34.7             &34.3                &  34.1    \\
$s/b$                          & &  &9.5              &  9.3               &  7.0      \\
$S/b$                          & &  &233.4            & 227.7              & 238.5     \\
$E/S$ [{\small MeV}]   $\!\!$  & &  &148.7            & 150.5              & 143.0       \\
\hline    
$\chi^2/$dof                   & &  &10/($21-3$)     &25/($21-3$)          &188/($21-2$) \\
\hline
 &  &  &     &        J.R./J.L. &  26.10.2002\\
\vspace*{0.2cm}
\end{tabular}
\end{table}

A global fit to the experimental data for RHIC-130 is given 
in table \ref{RHIChad}. We consider here 21  particle ratios. 
In some cases within the error other 
results from same and/or another collaboration are available, our selection is
somewhat subjective, but does not influence in  essential way the
general conclusions which follow. In the three last columns, the results 
for both chemical equilibrium (last column) and non-equilibrium fits
are seen. 

Next to the fitted ratios, we show in parenthesis the contribution to the 
error ($\chi^2$) for each entry.
We consider  statistical errors for the experimental results, 
since much of the systematic error should cancel in the particle ratios. However, 
we do not allow, when pion multiplicity is considered, for errors  smaller
than $\simeq 10$\%, which is our estimated error in the theoretical evaluation
of the pion yield due to incomplete understanding of the high mass
hadron resonances. Some of the experimental results are 
thus shown with a `theoretical' error. 
When such an enlargement  of the experimental error
is introduced, a dagger as superscript appears in the experimental 
data second column in  table \ref{RHIChad}. 

The high yields of hyperons
require significant (30-40\%) yield corrections for unresolved weak decays. 
Some experimental results are already corrected in this fashion:  the weak
cascading corrections were applied to the most recent $p$ and $\bar p$ 
results by the PHENIX collaboration \cite{Phe02}, and in the $\Xi/\Lambda$
and $\overline\Xi/\overline\Lambda$ ratio of the STAR collaboration we
use here \cite{Cas02}. However, some of the results
we consider are not yet corrected \cite{Phe01,Sta02}, and are indicated 
in the first column in table \ref{RHIChad} by a subscript $\Lambda$ or $\Xi$. 
 We present  two non-equilibrium fits, 
left  with 100\% $\Xi\to Y$ cascading acceptance  
and  with  40\% $Y\to N$ acceptance. Then, a  non-equilibrium 
fit with  40\% $\Xi\to Y$  and 40\% $Y\to N$, and in the last column, 
the chemical equilibrium fit with 40\% cascading. 

Below the fit  results, we show the statistical parameters  
which are related to each fit, and at the very bottom the $\chi^2$ of the fit.
We observe a considerable improvement in the statistical 
significance of the results of chemical non-equilibrium fits. 
The results, shown  in the table \ref{RHIChad}, are 
obtained minimizing in the space of 3 parameters: 
the chemical freeze-out temperature
$T$, and 2 chemical parameters $\lambda_q,\gamma_s$, the value of 
$\gamma_q$ is set at its maximal value, $\gamma_q^2=e^{m_\pi/T}$,
and the value of $\lambda_s$  is derived from the 
strangeness conservation constraint. Freeing these parameters
does not alter the results,  the  fit converges to local strangeness 
neutrality, within a few percent, and to full pion phase space
saturation.
 
We note that several particle yields
are not properly described in the last column of  table \ref{RHIChad}, 
and hence a large $\chi^2$ results in this chemical
equilibrium fit. However, 
on a logarithmic scale only results 
involving $\Lambda,\overline\Lambda$ would
be clearly visible as a discrepancy. The second and third last column 
show result of chemical non-equilibrium fits 
bracketing the $\Xi$-cascading, and yielding an excellent confidence level.

In the bottom of table~\ref{RHIChad}, we see
that the chemical non-equilibrium fit
specific strangeness content $s/b\simeq 9.5$ 
is nearly 40\% greater than the
chemical equilibrium result. This originates in 
$\gamma_s/\gamma_q\simeq 1.5$ ({\it i.e.,} $\gamma_s\simeq 2.5$). 
This specific yield   of strangeness and strangeness occupancy
 we measure in the phase space
after hadronization is consistent with the expected
 QGP properties before hadronization with nearly 
saturated strangeness phase space. The Wr\'oblewski ratio \cite{Wro85},
$W=2\langle s\bar s\rangle/(\langle u\bar u\rangle + \langle d\bar d\rangle)$,
is nearly proportional to $\gamma_s/\gamma_q$ and it  increases by 40\% 
at RHIC-130 compared to SPS. However, an analysis with 
$\gamma_s/\gamma_q\simeq 1$ will not report this enhancement \cite{Bec02}.

Our chemical nonequilibrium
 results yield a low freeze out temperature $T_h=140$--$142$\,MeV. This is 
considerably less than the temperature of phase transformation, $T_H$ which
might have been naively expected. This  range of hadronization
temperature now seen agrees better with the expectations we had 
upon consideration of the effect of the fast expansion  of QGP~\cite{Raf00}. 
We believe  that the wind of expending quarks and gluons adds
to the thermal pressure and the combined kinetic and thermal
pressure of exploding quark-gluon plasma can press out the confining vacuum
even at lower temperature \cite{Cso02}.

The results we find 
strongly favor chemical 
non-equilibrium description of the hadronization process.
A non-equilibrium chemical analysis of heavy-ion particle
yields offers profound insight into the physical properties of the
dense hadronic matter formed in the relativistic heavy-ion collisions
and allows to infer conditions prevailing in the deconfined phase. 
While the equilibrium Fermi model provides 
a first impression about the range of freeze-out temperatures, our
post-Fermi model resolves the `fine structure' of the hadronization yields.


\end{document}